# First-principles study of Carbz-PAHTDDT dye sensitizer and two Carbz-derived dyes for dye sensitized solar cells

Narges Mohammadi[*] and Feng Wang[*]

eChemistry Laboratory, Department of Chemistry & Biotechnology, School of Science,

Faculty of Science, Engineering and Technology

Swinburne University of Technology, Hawthorn, Melbourne, Victoria, 3122, Australia

* Corresponding authors: nmohammadi@swin.edu.au (Tel.: +61 3 9214 8785, N. Mohammadi).

fwang@swin.edu.au (Tel.: +61 3 9214 5065, Fax: +61 3 9214 5921, F. Wang).

**Abstract**

Two new carbazole-based organic dye sensitizers are designed and investigated *in silico*. These dyes are designed through chemical modifications of the π-conjugated bridge of a reference organic sensitizer known as Carbz-PAHTDDT (S9) dye. The aim of designing these dyes was to reduce the energy gap between their highest occupied molecular orbital (HOMO) and lowest unoccupied molecular orbital (LUMO) and to red-shift their absorption response compared to those of the reference S9 dye sensitizer. This reference dye has a reported promising efficiency when coupled with ferrocene-based electrolyte composition. To investigate geometric and electronic structure, density functional theory (DFT) and time-dependent DFT (TD-DFT) calculations were conducted on the new dyes as well as the reference dye. The present study indicated that the long-range correction to the theoretical model in the TD-DFT simulation is important to produce accurate absorption wavelengths. The theoretical studies have shown a reduced HOMO-LUMO gap and red-shifted absorption spectra for both of the new candidate dyes. In particular, the new S9-D1 dye is found to have significant reduced HOMO–LUMO energy gap, greater push–pull character and higher wavelengths of absorption when compared to the reference dye. Such findings suggest that the new dyes are promising and suitable for optoelectronic applications.

## 1. Introduction

With a booming in research effort to develop cost-effective renewable energy devices, dye sensitized solar cells (DSSC), also known as Grätzel cells [1] have been the topic of more than a thousand published papers just in 2010 [2]. Such considerable research interest in DSSC stems from the unique design and properties of this type of solar cells. Unlike conventional silicon-based solar cells, DSSC employs different components for light absorption, electron transport and hole transport. The low cost of the materials and processes involved in making DSSC is another advantage of this type of solar cells over the conventional silicon-based ones. However, the immense research effort to enhance efficiency of DSSC, which is still lower than that of silicon-based solar cells [3], has not been paired with adequate increase of the energy conversion efficiency of this device for commercialization.

The unique modular design of DSSC offers a wide range of possibilities to modify its components, such as dye sensitizer. In recent years, there has been an increasing interest in developing and employing new materials as a route to improve the conversion efficiency of dye sensitized solar cells. However, design and test of new materials for DSSCs have been dominated by synthesis procedures which are often costly and time-consuming [4]. As in the case of new dye sensitizer materials development, it is difficult for synthetic chemists to work out high-performance dyes with desirable properties prior to experiments on the assembled cell, without any support on the information of new dyes [5]. For example, the energy conversion efficiencies of the recently constructed DSSCs based on two chemically similar dyes were very different [6]. One is $\eta$=6.79% and the other is $\eta$=4.92%. And interestingly, these two dyes only differ slightly in their π-spacers: one takes thiophene ($\eta$=6.79%) and the other is thiazole ($\eta$=4.92%). Both spacers have a sulphur embedded in the pentagon ring, but the former contains two C=C bonds and the latter has one C=C bond and one C=N bond. Such structure and property relationship of the new dyes is hardly obtained without accurate quantum mechanical calculations.

In some cases, disappointing results from the most late-stage of the dye synthesis laboratories indicate an urgent need to understand the physical origin of dyes at molecular level, before experiments take place. To overcome this bottleneck in the development of new DCCSs with better efficiency, the state-of-art computational methods need to be utilised. Today, first-principle quantum chemical calculations are made available on supercomputing facilities which are accessible to more research groups. Such calculations are a reliable tool to design, study, and screen new materials prior to synthesis, in addition to probing the already existing dyes [7-13]. Computer-aided rational design of new dye sensitizers based on the systematic chemical modifications of the structure of already well-performing dyes has recently drawn the attention of several groups, including ours [14-23].

The recent breakthrough of the DSSC based on $Fc/Fc^+$ redox couple and the Carbz-PAHTDTT (S9) dye sensitizer [24], stimulated the present study with more theoretical insight to model and investigate this dye sensitizer, computationally. To the best of our knowledge, no detailed computational study on this dye is available. Computational study gives insight into the geometric and electronic structure of the dye sensitizer and serves as the starting point for rational design of new dyes with desirable properties such as improved spectral coverage. In addition to the study of the original Carbz-PAHTDTT (S9) dye, we design and investigate two new dyes through chemical modifications of the structure of this sensitizer. Our rational molecular design of the new dyes is aimed at red-shifting and broadening the absorption spectra of the S9 dye to utilise a greater fraction of the solar spectrum, which increases the photocurrent in the solar cell, and thus could increase the efficiency of the device [25-28].

It is also well-known that an important feature of organic push-pull dye sensitizers such as S9 dye is the intra-molecular charge transfer character (ICT). For example, the relationship between the ICT character of the molecules and their performance in DSSC has been studied by Edvinsson *et al.* [29] for the perylene-based sensitizers. It was found that the photocurrent and the overall solar-to-electrical energy conversion efficiency improve remarkably with increasing ICT character of the dyes. A later study of Tian and co-workers on triphenylamine dyes [30] suggests that an effective ICT has a positive effect on the performance of DSSC. It is known that the first hyperpolarizability ($β$) of a push-pull organic dye is associated with its ICT character [31-33]. As a result, this nonlinear optical (NLO) feature (i.e. $β$) of the original S9 dye and the new dyes will be probed in the present study.

2. Methods and computational details

The structure of the Carbz-PAHTDDT (S9) dye in three dimensional (3D) spaces was obtained through geometry optimizations in vacuum and in dichloromethane (DCM, $CH_2Cl_2$) solution, respectively. Density functional theory (DFT) based PBE0 hybrid density functionals [34] and polarized split-valence triple-zeta 6-311G(d) basis set, that is, the PBE0/6-311G(d) model, was employed in the calculations without any symmetry restrictions. No imaginary frequencies were found for the optimized structure, which ensures that optimized structure of S9 dye is a true minimum structure.

Several hybrid DFT functionals, namely, B3LYP, PBE0, BHandH and CAM-B3LYP with the same basis set (6-311G(d)) were employed for the calculations in DCM solution, in order to identify the most appropriate DFT functionals to calculate the S9 energy gap [24]. The conductor-like polarizable continuum model (C-PCM) [35,36] can effectively and accurately compute the influence of solute-solvent interactions on molecular energies, structures, and properties and is a good particularly model for large systems such as S9 dye. The model "has spread in the scientific community due to its accuracy and the relative simplicity of the expressions involved in the definition of the solvent reaction field" [36]. As a result, this model is employed to account for the solvent effects (i.e. DCM) on the absorption spectra and molecular energy levels in the present study.

The calculation of the hyperpolarizability is performed using PBE0/6-311G(d) model in vacuum. The present study investigates the total hyperpolarizability ($\beta_{tot}$) of the parent S9 dye and the new dyes. The tensor components of hyperpolarizability are obtained in the lower tetrahedral order, i.e. $\beta_{xxx}$, $\beta_{xxy}$, $\beta_{xyy}$, $\beta_{yyy}$, $\beta_{xxz}$, $\beta_{xyz}$, $\beta_{yyz}$, $\beta_{xzz}$, $\beta_{yzz}$, and $\beta_{zzz}$, respectively. The following equation gives the total hyperpolarizability [37,38]:

$$\beta_{tot} = [(\beta_{xxx} + \beta_{xyy} + \beta_{xzz})^2 + (\beta_{yyy} + \beta_{yzz} + \beta_{yxx})^2 + (\beta_{zzz} + \beta_{zxx} + \beta_{zyy})^2]^{1/2} \quad (1)$$

To accurately reproduce the experimental photo physical results, such as $\lambda_{max}$ of S9 dye sensitizer, several standard hybrid DFT functional (i.e., B3LYP [39], PBE0 [34] and BHandH [40]) and long-range corrected (LC) DFT functionals (such as CAM-B3LYP [41], ωB97XD [42] and LC-ωPBE [43-45]) have been included for the TDDFT calculations.

We have also investigated two *new* dyes which are designed by structural changes in the π-conjugated bridge of the reference Carbz-PAHTDDT (S9) dye. Both new dyes are computationally studied using the same method as that of the S9 dye study. All *ab initio* calculations were performed in Gaussian09 package [46].

**3. Results and discussion**

**3.1. Geometrical structures and design of the new dyes**

The backbone structure of the Carbz-PAHTDTT dye (S9) is given in Figure 1. This dye exhibits an electron-rich donor group (D), a π-conjugated bridge or linker and an acceptor moiety (A) as marked in the figure by three boxes. As a result, S9 is a D-π-A dye which is a common structure for organic dye sensitizers [4,47-52]. A two-carbazole-unit substituted triphenylamine group is employed as the electron donor unit (D) of the dye. It was previously shown that this donor structure suppresses the close π-stacked aggregation between the donor moieties of dye sensitizers adsorbed onto the surface of $TiO_2$ semiconductor [53]. Aggregation can result in intermolecular quenching and also leads to dye molecules which are not functionally attached to the semiconductor's surface and work like filters [54-57]. This phenomenon is known to be a detrimental factor of the efficiency for DSSC which should be avoided either by structural design or by employing co-adsorbents [58]. The non-coplanar structure of the electron donating moiety (D) can enhance thermal stability of dye sensitizer molecules by decreasing the contact between them. Thermal stability of dye sensitizer is an important factor for long term stability of functional solar cells [53].

The π-bridge (linker, the middle box in Figure 1) consists of five pentagon rings which are labelled as I, II, III, IV and V in Figure 1. A dithienothiophene (DTT) unit forms central part of the π-conjugated bridge of the S9 dye. This moiety leads to a better stability of the dye sensitizer in high polarity electrolytes used in DSSC. To provide additional double conjugation into the linker moiety [59], two hexanyl ($C_6H_{13}$) chain-substituted thiophene rings (i.e. 3-hexylthiophene or rings I and V in Figure 1) exist in the π-conjugated bridge of the S9 dye which can form either trans or cis isomers. A *cis*-S9 is formed when both of the hexanyl chains ($C_6H_{13}$) are in the same side of the π-bridge, or a *trans*-S9 isomer is formed if the hexanyl chain ($C_6H_{13}$) groups locate in the different sides of the π-bridge. The long hexanyl chains suppress the aggregation of the dye molecules, and also enable longer electron life time ($\tau$) [60]. The present calculations indicate that the *cis*-S9 isomer possesses a total energy of approximately 4.6 kJ·mol$^{-1}$ less than the total energy of the trans conformer, indicating that the S9 dye slightly favours the cis conformation. Therefore, only *cis* conformation will be discussed here (results and discussion of the trans conformation are given in the Online Resource 1). On the acceptor side of this dye (A), the conventional acceptor moiety is employed which contains the cyano group as an electron withdrawing group and the carboxyl group as an anchoring unit to attach the dye onto the $TiO_2$ semiconductor.

As learned from our previous study on rational design of D-π-A dyes [20], all three moieties of a dye, i.e., the donor, the π-bridge and the acceptor can be modified to produce new dyes. For the purpose of the rational design in this study, the π-bridge of the S9 dye has been modified to produce a pair of new dyes. In the π-conjugated bridge of this dye, a dithienothiophene unit (DTT) is employed. Kwon *et al.* who synthesized S9 dye, have also reported another DTT-based dye sensitizer (DAHTDTT 13) with a similar structure to S9 dye which only differs in its D group [59]. The absorption spectra of these two dye sensitizers are very similar for the visible portion of the spectrum, i.e., λ>400 nm. As a result, in the present study, instead of making changes in the D-group and A-group, the linker (i.e., the π-bridge) of the S9 dye is modified.

An aim of the design of the new dyes is to extend the absorption spectra to near infrared (NIR) region by reducing the HOMO-LUMO energy gap of the dye sensitizer. As a result, two new derivatives dyes (S9-D1 and S9-D2) are designed from the original (*cis*-) Carbz-PAHTDTT (S9) dye through the modification of the π-bridge linker. In S9-D1, the $X_1$ and $X_2$ groups in S9 dye were replaced by the N= groups, respectively, but in S9-D2, the $X_1$ and $X_2$ groups were substituted by the -NH groups, respectively. (Please refer to Figure S1 the in Online Resource 1 for the obtained π-bridges of the optimized structures, Carbz-PAHTDTT (S9), S9-D1 and S9-D2.)

Table 1 lists the important molecular properties of the Carbz-PAHTDTT (S9) dye (*cis* isomer) and the new S9-D1 and S9-D2 dyes obtained in vacuum. As all the dyes are either new dyes (S9-D1 and S9-D2) or recently synthesised dye (S9), only very limited information is available for comparison. However, the models such as the PBE0/6-311G(d) model have been shown to be reliable in previous studies [20,34,61]. The π-conjugated bridge length ($L_\pi$) of the D-π-A dye, defined as the direct distance between $C_{(43)}$ and $C_{(61)}$, is calculated at 17.14 Å for S9 dye which is shortened in S9-D1 (16.33 Å) and S9-D2 (16.52 Å) as the N atoms (S9-D1 and S9-D2) have smaller radius than S atoms (S9), which is also reflected by the calculated molecular sizes (i.e., the electronic spatial extent $<R^2>$) of the dyes. The dipole moments (μ) of the S9-D2 dye (μ=5.12 Debye) exhibit very similar values to the original S9 dye (μ=5.10 Debye), whereas the dipole moment of S9-D1 (6.72 Debye) exhibits larger changes.

### 3.2. Electronic and optical properties

The experimental energy values for the HOMO, LUMO and HOMO-LUMO gap of the Carbz-PAHTDTT (*cis*-S9) dye in $CH_2Cl_2$ (DCM) solution are estimated at -5.08 eV, -2.97 eV and 2.11 eV, respectively [24] (refer to Table S1 in Online Resource 1 for more details). In the present study, several functionals are employed to calculate the frontier molecular orbital energies of S9, S9-D1, and S9-D2 dyes in DCM solution. Calculated values by different functionals are given in Table S2 in Online Resource 1. Values obtained from B3LYP functionals for the reference S9 dye are calculated at -5.08 eV, -2.99 eV and 2.08 eV for HOMO, LUMO and HOMO-LUMO gap, respectively. These calculated values are in excellent agreement with the experiment by less than 0.02 eV deviations from the experimental values. Thus CPCM-B3LYP/6-311G(d)//CPCM-PBE0/6-311G(d) model is selected to predict the energy levels of new dyes S9-D1 and S9-D2 and to construct the following molecular energy levels graph and isodensity plots.

Figure 2 compares the calculated frontier molecular orbital energy levels of the S9 dye and the new dyes S9-D1 and S9-D2 in DCM solution, focusing on the HOMO–LUMO energy gap. As seen in this figure, the HOMO-LUMO gap of the dyes reduces from 2.08 eV in S9 to 1.66 eV in S9-D1 and 1.88 eV in S9-D2. However, the HOMO-LUMO gap reductions in S9-D1 and S9-D2 are very different: in S9-D1, the most significant change is the reduction of the LUMO energy, from -2.99 eV in S9 to -3.66 eV in S9-D1, although the energy of the HOMO of S9-D1 also exhibits a small reduction, from -5.08 eV (S9) to -5.32 eV (S9-D1). On the other hand, the HOMO-LUMO gap reduction in S9-D2 dye is achieved by shifting up the energy level of the HOMO, from -5.08 eV to -4.79 eV, whereas the energy of the LUMO almost remains the same as -2.99 eV in S9 and -2.91 eV in S9-D2. The HOMO-LUMO energy gap reductions of the new dyes in Figure 2 are also seen in the corresponding orbitals of the dyes. When the orbitals are more localised, the energies increase, whereas when the orbitals are more delocalised, the energies decrease.

The new dye S9-D1 has similar HOMO distribution to the reference structure but with more contribution from the carbazole-units in the donor moiety and the second hexanyl chain-substituted thiophene ring (i.e. ring V) in the linker moiety. However, the LUMO of S9-D1 is extended into a phenyl group in the donor moiety of the dye structure and a small density decrement is observed on the carboxyl group (A section) which might lead to a weaker electron coupling with semiconductor surface compared to the reference S9 structure. Also, S9-D2 dye shows very similar HOMO and LUMO distribution to the reference S9 dye with the HOMO being slightly shifted from the donor section toward the π-conjugated bridge.

The tensor components of the first order hyperpolarizability ($\beta$) of S9, S9-D1 and S9-D2 dyes are given in Table S3 and the Cartesian axes are displayed in Figure S2 of Online Resource 1. The total hyperpolarizabilities, $\beta_{tot}$, are calculated as 805 esu for S9, 2190 esu for S9-D1 and 856 esu for S9-D2. The $\beta_{tot}$ values of both new dyes, S9-D1 and S9-D2, are significantly higher than the reference dye. However, a dramatic increase of almost 2.7 times is calculated for the magnitude of the $\beta_{tot}$ of the new dye S9-D1. The molecular hyperpolarizabilities of the studied dyes are dominated by $\beta_{xxx}$ (Please see Table S3). The magnitude of the highest $\beta$ tensor component (i.e. $\beta_{xxx}$) is almost comparable for S9 and S9-D2 dyes, whereas that of the S9-D1 is significantly higher. For all three dyes calculated $x$ component of the total hyperpolarizability, $\beta_x = (\beta_{xxx} + \beta_{xyy} + \beta_{xzz})^2$ is dramatically higher in magnitude than the $y$ component of the total hyperpolarizability, $\beta_y = (\beta_{yyy} + \beta_{yzz} + \beta_{yxx})^2$, which is significantly higher in magnitude than the *z* component of the total hyperpolarizability, $\beta_z = (\beta_{zzz} + \beta_{zxx} + \beta_{zyy})^2$, i.e., , $\beta_x > \beta_y > \beta_z$.

The enhanced NLO properties of the S9-D1 are attributed to the planarity of the π-conjugated bridge of this dye compared to both S9 and S9-D2 dyes. The non-planarity in S9 and S9-D2 is resulted from the sigma (σ) bond between the pentagon rings I and II, which reduces the overlap of the interacting orbitals and consequently will reduce the ICT from donor to acceptor ends of these dyes. On the other hand, because of the chemical modifications made in S9-D2 dye, the aforementioned sigma bond becomes a double bond in S9-D2, which prevents the free rotation around this bond. As a result, the overlap of the interacting orbitals is enhanced in S9-D2 and its ICT character is increased which will be manifested through the dramatic increase of the hyperpolarizability of this sensitizer. Our results together with other studies show that HOMO-LUMO gap is a critical factor in determining the $\beta_{tot}$ value [37,31,33]. That is, the highest $\beta_{tot}$ is observed for the molecule with the smallest energy gap (i.e. S9-D2). The higher $\beta_{tot}$ values of S9-D1 suggests that this molecule possesses better push-pull properties which can enhance the electron charge transfer capability of this dye sensitizer and consequently the efficiency (*η*) of the solar cell.

**3.3. Excitation energies and UV-Vis spectra**

The experimental absorption spectrum of the S9 dye was measured in the dichloromethane (DCM) solution [24]. The three main absorption bands in the UV-Vis spectral region of 300-800 nm of the S9 dye are reported at $\lambda_1$=491 nm, $\lambda_2$=426 nm, and $\lambda_3$=330 nm [24]. Table 2 compares the experimental measurement with theoretical results using TDDFT with respect to different DFT models indicated in the previous section for the original S9 dye. To assess the overall performance of TDDFT functionals with reference to the experimental values in this table, the mean absolute error (MAE) criterion is employed as,

$$\text{MAE} = \frac{1}{n} \sum_{i=1}^{n} \left| \lambda_i^{calc.} - \lambda_i^{expt.} \right| \quad , (n = 3) \qquad (2)$$

The MAEs of the DFT functionals are given in the last row of Table 2. In the table, the CAM-B3LYP model and the BHandH are compatible with MAEs being 14 nm and 18 nm, respectively. Next comes the ωB97XD (MAE=23 nm) and the LC-ωPBE (MAE=52 nm) models. The PBE0 (MAE=108 nm) and B3LYP (MAE=149 nm) models exhibit the least accurate performance on prediction of the spectral line positions. Three long-range (LC) corrected DFT functionals, namely CAM-B3LYP, ωB97XD and LC-ωPBE show a good general performance in reproducing the experimental main bands. Non-LC hybrid functionals are not usually suitable and accurate at large distances for electron excitations to high orbitals. The present results are in good agreement with other studies [62,63,13,7,64,65], suggesting that long-range corrected CAM-B3LYP functional can be a good model in the TDDFT calculations of large-sized organic dye molecules with a spatially extended π system, as the charge transfer transitions take place through space in such dyes.

The results in Table 2 also suggest that without long-range corrections, the inclusion of the Hartree-Fock (HF) exchange energy is important to reproduce the major band, $\lambda_1$. The TD-BHandH DFT model gives the major absorption peaks of the S9 dye the most accurate results and the TD-B3LYP DFT model produces the least accurate results in the table. For example, the TD-BHandH DFT model almost reproduces the spectral line position of the major absorption band at $\lambda_1$= 490 nm with respect to the experiment at $\lambda_1$=491 nm. This model also closely reproduces the other minor bands at $\lambda_2$=402 nm (expt. 426 nm) and $\lambda_2$=360 nm (expt. 330 nm). Without sufficient inclusion of the exchange energy in the DFT functionals, the B3LYP and PBE0 hybrid functionals are unable to produce spectral band positions with sufficient accuracy as seen previously [66,67].

Table 2 also collects the effect of modifications on shifting the spectral peaks. For S9-D1, a remarkable bathochromic shift (i.e. to the longer wavelengths or red-shift) of 172 nm on $\lambda_1$ compared to the reference S9 dye is observed. This band is mainly composed of an excitation transition from HOMO → LUMO both for the reference S9 dye and for the new S9-D1 structure (refer to Table S4 in Online Resource 1 for detailed assignment of the transitions). Figure 3 reports the simulated UV-vis spectra of the three dyes, S9, S9-D1 and S9-D2. As discussed earlier, the energy gap between HOMO and LUMO of S9-D1 is significantly reduced compared to that of S9 dye, which in turn results in the red-shift of $\lambda_1$ as seen in this figure. Very large red-shift of $\lambda_2$ and $\lambda_3$ are also observed for S9-D1 compared to the S9 dye. The $\lambda_1$ band of the new dye S9-D1 indeed outperforms the original Carbz-PAHTDTT (S9) dye with not only a significant preferred spectral shift on the position of this band, but also this band covers a broader region with nearly doubled full width at half maximum (FWHM) than the original S9 dye. The significant bathochromic shift and broadening of the UV-Vis spectra of S9-D1 structure indicates its enhanced light harvesting capability which is an important criterion for a well-performing dye sensitizer employed in DSSC. In the S9-D2 dye, the $\lambda_1$ and $\lambda_3$ spectral bands both show a preferred bathochromic shift of 44 nm compared to the S9 dye. However, an unwanted hypsochromic shift (i.e. to the shorter wavelengths or blue-shift) of -32 nm on the $\lambda_2$ spectral band was calculated for this (S9-D2) dye.

## 4. Conclusions

The recently published Carbz-PAHTDTT (S9) organic dye sensitizer was studied theoretically using a number of DFT models in vacuum and in DCM solution. Good agreement with experimental results indicate that the B3LYP/6-311G(d)//PBE0/6-311G(d) model in solution can be utilized to predict the energy levels of HOMO and LUMO of new similar dyes which are designed and studied *in silico*. However, to produce good agreement with the experimental UV-Vis spectra of the S9 dye, long-range corrected functionals, such as CAM-B3LYP and half-and-half functionals such as BHandH functional need to be considered.

In the present study, a pair of new dyes, S9-D1 and S9-D2, was designed through chemical modifications of the π-conjugated bridge of the reference S9 dye. These new dyes showed a reduced HOMO-LUMO energy gap compared to that of the S9 dye, through lowering the LUMO energy (S9-D1) or lifting up the HOMO energy (S9-D2). A significant red-shifting and broadening of the resulted absorption spectrum of the S9-D1 is achieved. The present study explored a useful direction of rational design for new dyes in DSSC.

## Acknowledgements

NM would like to thank Swinburne University Vice-chancellor's Postgraduate Award. NM and FW thank Victorian Partnership for Advanced Computing (VPAC) and Swinburne University Supercomputing Facilities for computer resources.

**Table 1:** The selected bond length, dihedrals, π-lengths[a] and dipole moment of the S9, S9-D1 and S9-D2 dyes*.

|  | S9 | S9-D1 | S9-D2 |
|---|---|---|---|
| $L_\pi$ [a] (Å) | 17.14 | 16.33 | 16.52 |
| $C_{48}$-$C_{49}$ (Å) | 1.44 | 1.37 | 1.44 |
| $X_1$-$C_{48}$-$C_{49}$-$X_2$ (°) | -144.91 | -179.09 | -157.07 |
| $X_2$-$C_{52}$-$C_{53}$-$S_4$ (°) | 0.93 | -0.79 | -0.10 |
| $S_4$-$C_{56}$-$C_{57}$-$S_5$ (°) | 150.10 | 157.57 | 148.19 |
| $<R^2>$ (a.u) | 275867.82 | 256546.52 | 260606.50 |
| µ (Debye) | 5.10 | 6.72 | 5.12 |

*Optimized at PBE0/6-311G(d) level.

(a) Direct distance of $C_{(43)}$-$C_{(61)}$.

**Table 2: The most intense absorption peaks of the S9, S9-D1 and S9-D2 dyes.**

| Method[a] | Carbz-PAHTDTT (S9) | | | | | | | S9-D1 | | S9-D2 | |
|---|---|---|---|---|---|---|---|---|---|---|---|
| | TD-B3LYP | TD-PBE0 | TD-LC-ωPBE | TD-ωB97XD | TD-CAM-B3LYP | TD-BHandH | Exp.[b] | TD-BHandH | $\Delta\lambda$[c] | TD-BHandH | $\Delta\lambda$[d] |
| $\lambda_1$(nm) | 668 | 611 | 420 | 460 | 479 | 490 | 491 | 662 | 172 | 535 | 45 |
| $\lambda_2$(nm) | 540 | 498 | 365 | 391 | 401 | 402 | 426 | 528 | 126 | 394 | -8 |
| $\lambda_3$(nm) | 488 | 463 | 306 | 325 | 335 | 360 | 330 | 440 | 80 | 374 | 14 |
| MAE[e] | 149 | 108 | 52 | 23 | 14 | 18 | | | | | |

(a) All TDDFT calculations are performed in DCM solution using CPCM solvation model on geometries optimized at CPCM -PBE0/6-311G(d).

(b) See supplementary information of [24].

(c) $\Delta\lambda = \lambda(S9\text{-}D1) - \lambda(S9)$, method= TD-BHandH.

(d) $\Delta\lambda = \lambda(S9\text{-}D2) - \lambda(S9)$, method= TD-BHandH.

(e) $\text{MAE} = \frac{1}{n} \sum_{i=1}^{n} \left| \lambda_i^{calc.} - \lambda_i^{expt.} \right|$ , (n = 3)

**Figure Captions**

**Fig. 1:** Sketch of the reference S9 dye and the new dyes S9-D1 and S9-D2.

**Fig. 2:** Calculated frontier MO energy levels using B3LYP/6-311G(d) // PBE0/6-311G(d) model in DCM solution and isodensity surfaces of the HOMO and LUMO of S9, S9-D1 and S9-D2 dyes.

**Fig. 3:** The simulated UV-Vis spectra of three dyes, S9, S9-D1 and S9-D2 using TD-BHandH/6-311G(d) model in DCM solution. A full width at half-maximum (fwhm) 3000 cm$^{-1}$ of the Gaussian curves is used to convolute the spectrum.

**Fig.1**

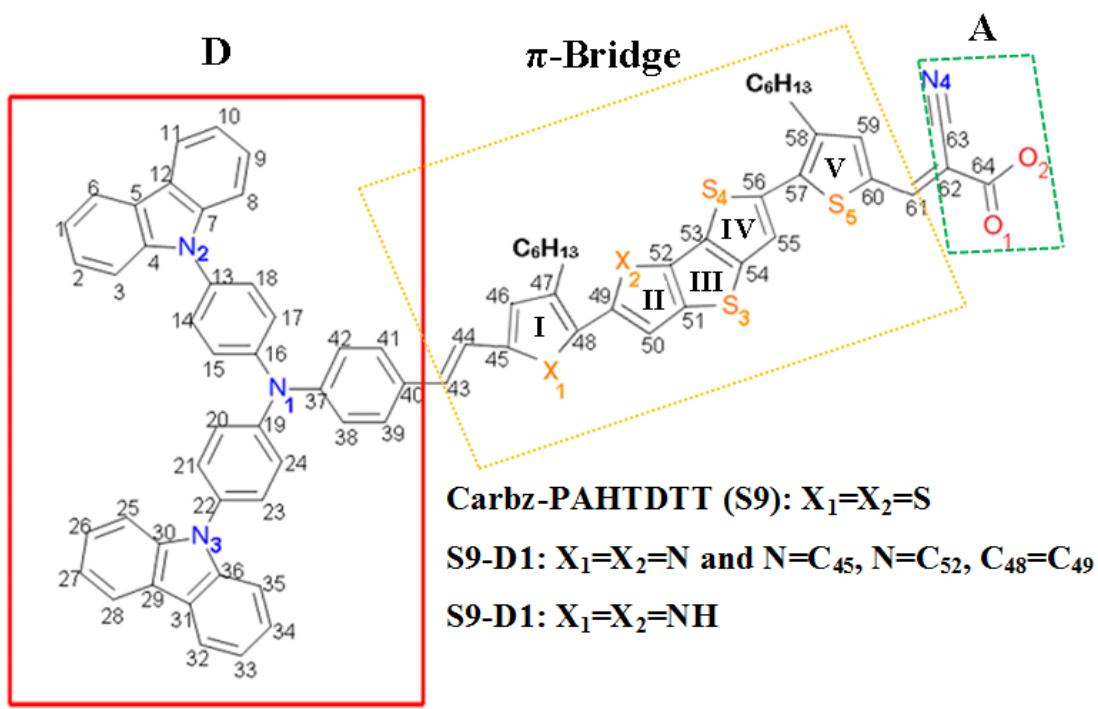

**Fig. 2**

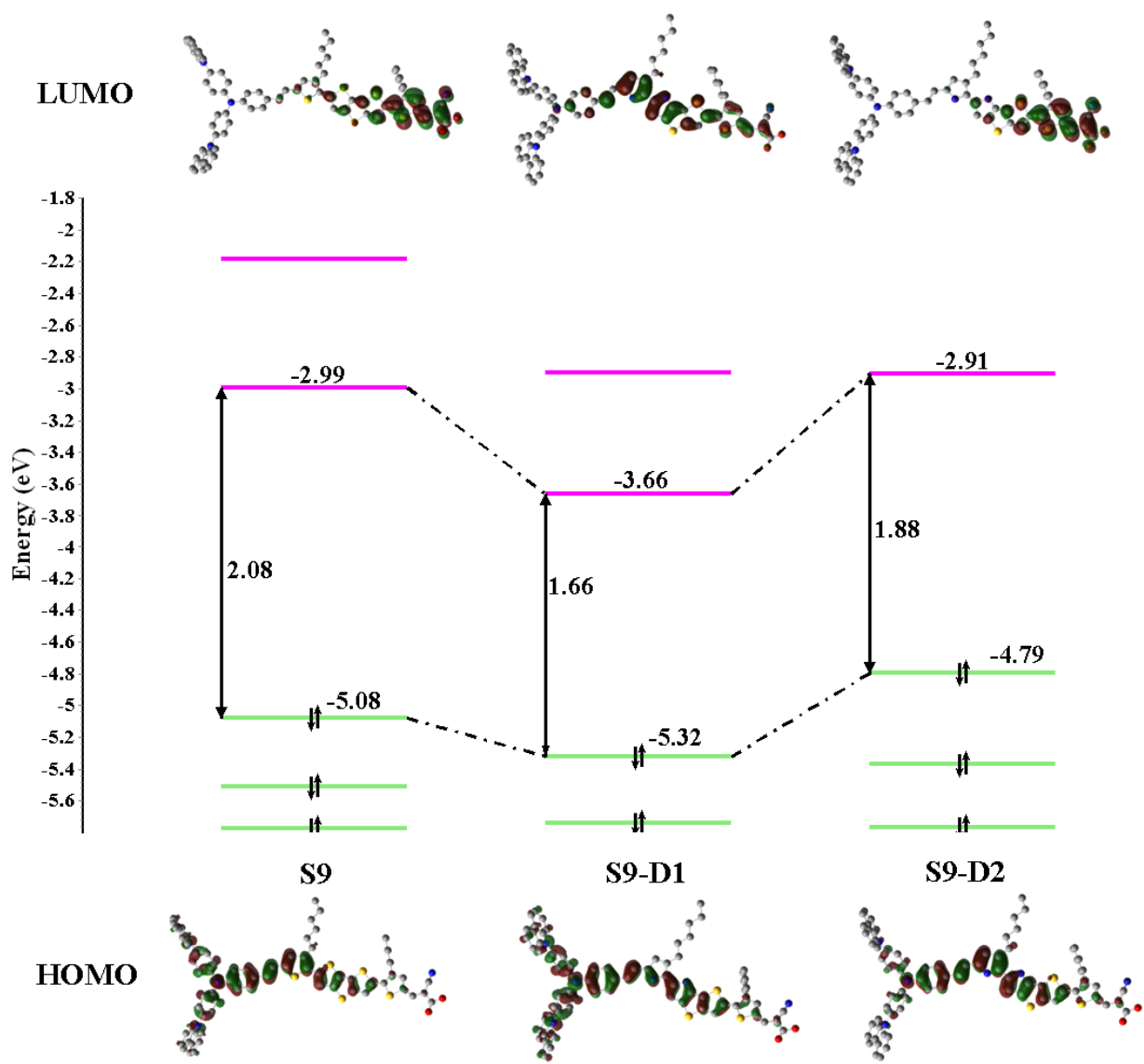

**Fig. 3**

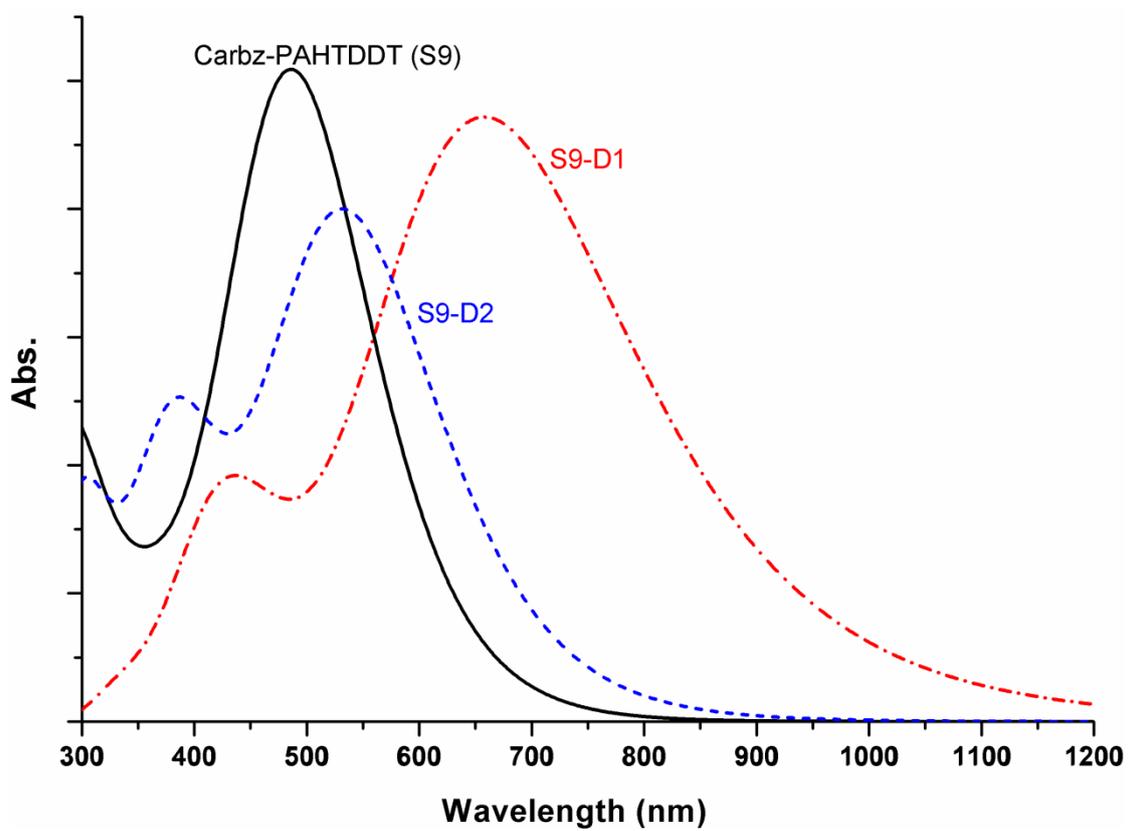

**First-principles study of Carbz-PAHTDDT dye sensitizer and two Carbz-derived dyes for dye sensitized solar cells**

Narges Mohammadi[*] and Feng Wang[*]

Department of Chemistry & Biotechnology, School of Science,
Faculty of Science, Engineering and Technology
Swinburne University of Technology, Hawthorn, Melbourne, Victoria, 3122, Australia

* Corresponding authors:
nmohammadi@swin.edu.au (Tel.: +61 3 9214 8785, N. Mohammadi).
fwang@swin.edu.au (Tel.: +61 3 9214 5065, Fax: +61 3 9214 5921, F. Wang).

**NBO Charges:**

To analyse the charge population of the dye sensitizer, natural bond orbital (NBO) analysis was performed on the optimized structure in vacuum using the NBO 3.1 program [1] embedded into Gaussian09 package. The computations of the NBO were carried out using the PBE0/6-311G(d) model in vacuum.

Atomic charges according to the natural bond orbital (NBO) scheme of the π-conjugated bridges of the dyes are given in Figure S1(a)-(c) for three dyes S9, S9-D1 and S9-D2, respectively. It is not a surprise that the atomic charges change more apparently at the positions local to the $X_1$ and $X_2$ atoms in the new dyes, whereas only small changes in atoms away from $X_1$ and $X_2$ are observed. For example, atoms $C_{(45)}$ and $C_{(48)}$ directly bond with $X_1$, whereas $C_{(49)}$ and $C_{(52)}$ directly bond with $X_2$. The atomic charges for these sites in S9 are negative, i.e., $X_1$=S: -0.185 a.u. for $C_{(45)}$ and -0.213 a.u. for $C_{(48)}$; and $X_2$=S: -0.186 a.u. for $C_{(49)}$ and -0.246 a.u. for $C_{(52)}$. However, the atomic charges in these sites in the derivatives, S9-D1 and S9-D2, switch their signs from being negative in S9 to being positive in S9-D1 and S9-D2, regardless the substituted groups are electron donating (-NH-) or electron withdrawing (-N-) groups.

The total NBO charges of the π-conjugated bridge (linker) in the D-π-A dyes can be either negative or positive. However, the overall net charge for the donor section (D) of the dyes is always positive, whereas the acceptor section (A) of the dyes is always negative. Although in the new dyes similar trend exist in the changes of the NBO atomic charges locally, the total NBO charges over the π-conjugated bridge (linker) of the dyes are not the same. Such the total charges over the π-conjugated bridge (linker) are calculated at +0.062 a.u., -0.029 a.u. and +0.118 a.u., respectively, for S9, S9-D1 and S9-D2. The fact that the original S9 dye and the new dye S9-D2 possess positive charges of the linker suggests that π-conjugated bridges of these dyes exhibit electron-donating character. On the contrary, the negatively charged linker of the S9-D1 dye shows that the chemical modifications with the –NH- group alter the electron-donating character of the linker in the original dye (S9) to an electron-withdrawing character of the linker in the S9-D1 dye.

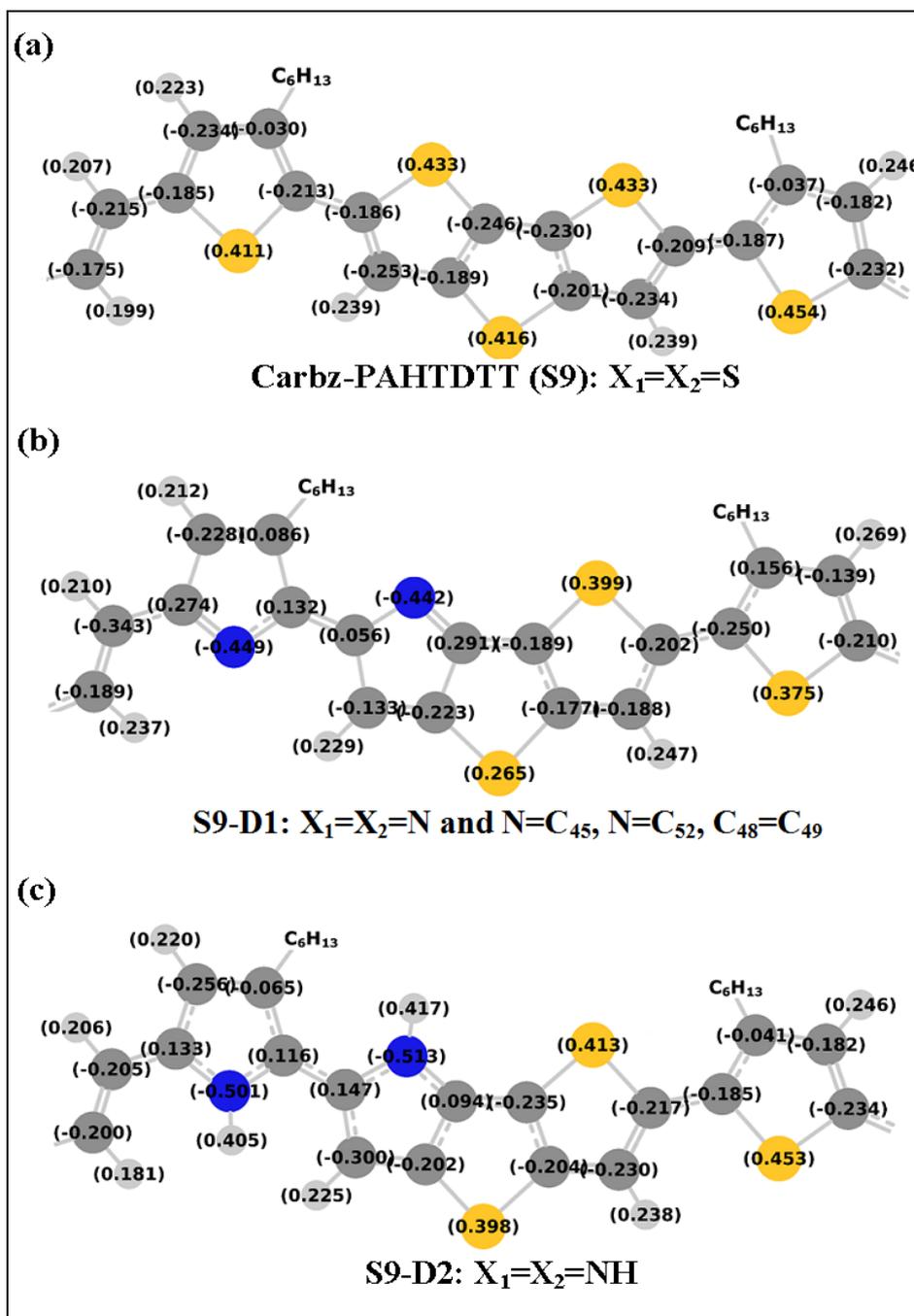

Fig S1: The structure of the π-conjugated bridge of S9, S9-D1 and S9-D2 dyes showing NBO charge of atoms in the linker. Note that hexanyl chains are not included.

**Table S1: The experimental values from the electrochemical measurements in dichloromethane solution**

| HOMO$_{expt}$ (eV)[a] | LUMO$_{expt}$ (eV)[a] | $E_{ox}$ (V) vs (NHE)[b] | $E_{0-0}$ (V) vs (Abs/Em)[b] | $E_{ox}$- $E_{0-0}$ (V) vs (NHE)[b] |
|---|---|---|---|---|
| -5.08 | -2.97 | 0.91 | 2.11 | -1.20 |

[a.] Experimental HOMO and LUMO are estimated as:

"HOMO = -($E_{onset}$ vs $Fc^+/Fc$ -4.8 eV), LUMO = HOMO + $E_{0-0}$." As described in (Table 1, P.4090) of Ref [2].

[b.] Values taken from the supplementary information file (Table S3, P.17) of Ref [3].

The frontier molecular orbitals energy levels with different levels of theory (B3LYP, PBE0, BHandH and CAM-B3LYP) are calculated. Results are listed in Table S1. Values listed in this table suggest that the same trend exists for the HOMO-LUMO energy gap ($\Delta$) of studied dyes regardless of the functionals employed to calculate them. That is, $\Delta_{S9-D1} < \Delta_{S9-D2} < \Delta_{S9}$. Another result emerged from this table is that for all three dyes, the $\Delta$ values calculated by the CAM-B3LYP functional are the highest, followed by those of BHandH functional and then by PBE0 functional and finally by B3LYP functional.

**Table S2: Energy levels of HOMO, LUMO and HOMO-LUMO gap calculated by different functionals.**

| Structure | Model[a] | HOMO (eV) | LUMO (eV) | GAP (eV) |
|---|---|---|---|---|
| | **Exp.** | **-5.08** | **-2.97** | **2.11** |
| **Carbz-PAHTDTT (S9)** | CPCM –B3LYP/6-311G* | -5.08 | -2.99 | 2.08 |
| | CPCM -PBE0/6-311G* | -5.31 | -2.94 | 2.37 |
| | CPCM - BHandH/6-311G* | -5.90 | -2.17 | 3.72 |
| | CPCM – CAM-B3LYP/6-311G* | -6.24 | -2.01 | 4.23 |
| **S9-D1** | CPCM –B3LYP/6-311G* | -5.32 | -3.66 | 1.66 |
| | CPCM -PBE0/6-311G* | -5.54 | -3.65 | 1.89 |
| | CPCM - BHandH/6-311G* | -6.12 | -2.98 | 3.14 |
| | CPCM – CAM-B3LYP/6-311G* | -6.44 | -2.83 | 3.61 |
| **S9-SD2** | CPCM –B3LYP/6-311G* | -4.79 | -2.91 | 1.88 |
| | CPCM -PBE0/6-311G* | -5.01 | -2.86 | 2.15 |
| | CPCM - BHandH/6-311G* | -5.53 | -2.10 | 3.43 |
| | CPCM – CAM-B3LYP/6-311G* | -5.89 | -1.94 | 3.95 |

(a) All single point energy calculations are performed on geometries optimized at CPCM - PBE0/6-311G*.

**Table S3: First hyperpolarizability tensor components (in a.u) and total hyperpolarizability (in esu).**

| Structure | $\beta_{xxx}$ | $\beta_{xxy}$ | $\beta_{xyy}$ | $\beta_{yyy}$ | $\beta_{xxz}$ | $\beta_{xyz}$ | $\beta_{yyz}$ | $\beta_{xzz}$ | $\beta_{yzz}$ | $\beta_{zzz}$ | $\beta_{tot}*10^{-30}$ (esu)[a] |
|---|---|---|---|---|---|---|---|---|---|---|---|
| **S9(*cis*)** | -89930.67 | 11813.59 | -1907.39 | 1134.09 | 9127.10 | -2829.37 | 695.71 | -1302.36 | 506.73 | 465.50 | 804 |
| **S9-D1** | 252260.26 | -8121.80 | 793.58 | -590.04 | -6977.06 | 178.46 | -69.77 | 442.54 | 221.70 | -315.85 | 2190 |
| **S9-D2** | 94044.55 | 17273.00 | 1781.94 | 818.41 | -18572.64 | -4008.86 | -624.57 | 3242.25 | 773.48 | -948.92 | 856 |

(a) $\beta_{tot}$ is converted from atomic unit (a.u) into electrostatic unit (1 a.u=8.6393 * $10^{-33}$ esu).

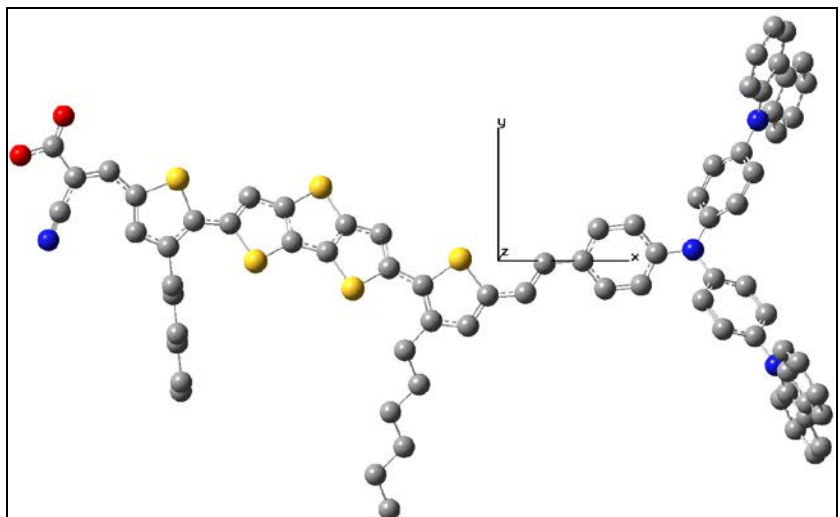

Figure S2: The optimized structure of S9 and the Cartesian axes.

**Table S4:** Calculated excited energy (in nm), transition configuration, and oscillator strengths (f) for the 3 most intense peaks of S9, S9-D1 and S9-D2 dyes in DCM solution.

| | Carbz-PAHTDTT(*cis*-S9) | | | | | | | (*trans*-S9) | S9-D1 | S9-D2 |
|---|---|---|---|---|---|---|---|---|---|---|
| Method | TD-B3LYP | TD-PBE0 | TD-LC-ωPBE | TD-ωB97XD | TD-CAM-B3LYP | TD-BHandH | Exp. | TD-BHandH | TD-BHandH | TD-BHandH |
| $\lambda_1$ | 668 (1.12)<br>H→L (97%)<br>H-1→L (2%) | 611 (1.48)<br>H→L (90%)<br>H-1→L (6%)<br>H→L+1 (3%) | 420 (2.83)<br>H-1→L (29%)<br>H→L (25%)<br>H→L+1 (12%)<br>H-3→L (10%)<br>H-6→L (6%)<br>H-6→L+1 (3%)<br>H-1→L+2 (2%) | 460 (2.87)<br>H→L (32%)<br>H-1→L (28%)<br>H→L+1 (15%)<br>H-3→L (9%)<br>H-6→L (4%)<br>H-6→L+1 (2%) | 479 (2.74)<br>H→L (40%)<br>H-1→L (28%)<br>H→L+1 (13%)<br>H-3→L (8%)<br>H-6→L (3%) | 490 (2.71)<br>H→L (52%)<br>H-1→L (23%)<br>H→L+1 (12%)<br>H-3→L (6%) | 491 (3.0) | 467 (2.51)<br>H→L (41%)<br>H-1→L (34%)<br>H-3→L (9%)<br>H→L+1 (8%)<br>H-6→L (2%) | 662 (2.55)<br>H→L (78%)<br>H-1→L (14%)<br>H-7→L (2%) | 535 (2.17)<br>H→L (76%)<br>H-1→L (9%)<br>H-3→L (5%)<br>H→L+1 (5%) |
| $\lambda_2$ | 540 (0.80)<br>H-1→L (91%)<br>H→L (3%)<br>H→L+1 (4%) | 498 (0.71)<br>H-1→L (82%)<br>H→L (8%)<br>H-3→L (5%)<br>H→L+1 (4%) | 365 (0.43)<br>H→L+1 (48%)<br>H-6→L (10%)<br>H-3→L (10%)<br>H-1→L (6%)<br>H→L+2 (4%)<br>H-8→L (3%)<br>H-1→L+1 (3%)<br>H-3→L+1 (2%)<br>H-1→L+7 (2%) | 391 (0.40)<br>H→L+1 (53%)<br>H-3→L (11%)<br>H-6→L (9%)<br>H-1→L (9%)<br>H→L+2 (4%)<br>H-8→L (2%)<br>H-1→L+7 (2%) | 401 (0.40)<br>H→L+1 (57%)<br>H-3→LO (10%) H-1→LO (11%)<br>H-6→L (7%)<br>H→L+2 (4%) | 402 (0.39)<br>H→L+1 (60%)<br>H-1→L (14%)<br>H-3→L (10%)<br>H-6→L (5%)<br>H→L+2 (3%) | 426 (2.7) | 389 (0.75)<br>H→L+1 (61%)<br>H-1→L (11%)<br>H-3→L (9%)<br>H→L+2 (4%)<br>H-6→L (4%)<br>H-3→L+1 (2%) | 528 (0.13)<br>H-3→L (45%)<br>H-6→L (26%)<br>H-1→L (17%)<br>H→L (4%) | 394 (0.84)<br>H→L+1 (72%)<br>H-1→L (7%)<br>H-3→L (6%)<br>H-6→L (2%) |
| $\lambda_3$ | 488 (1.10)<br>H→L+1 (88%)<br>H-3→L (3%)<br>H-1→L (5%) | 463 (0.94)<br>H→L+1 (89%)<br>H-1→L (6%) | 306 (0.29)<br>H→L+2 (27%)<br>H-1→L+1 (21%)<br>H-8→L (12%)<br>H-3→L+1 (7%)<br>H-6→L (6%)<br>H→L+7 (5%)<br>H-1→L+2 (2%) | 325 (0.29)<br>H-1→L+1 (26%)<br>H→L+2 (26%)<br>H-8→L (11%)<br>H-6→L (9%)<br>H-3→L+1 (6%)<br>H→L+7 (5%) | 335 (0.33)<br>H→L (38%)<br>H-6→L (15%)<br>H-3→L (11%)<br>H→L+2 (10%)<br>H-8→L (8%)<br>H→L+1 (9%)<br>H-1→L (3%) | 360 (0.29)<br>H→L (44%)<br>H-1→L (20%)<br>H→L+1 (13%)<br>H-3→L (9%)<br>H-6→L (5%)<br>H-1→L+1 (3%) | 330 (2.5) | 356 (0.38)<br>H→L (54%)<br>H-1→L (13%)<br>H→L+1 (13%)<br>H-3→L (8%)<br>H-6→L (4%)<br>H→L+2 (2%) | 440 (0.21)<br>H-1→L (45%)<br>H-6→L (30%)<br>H→L (8%)<br>H-8→L (4%)<br>H→L+1 (4%)<br>H-7→L+1 (2%)<br>H-3→L+1 (2%) | 374 (0.52)<br>H-1→L (29%)<br>H-3→L (22%)<br>H→L (22%)<br>H→L+1 (11%)<br>H-6→L (10%)<br>H-8→L (4%) |

**Results and discussions of the *trans-S9* conformation:**

Since two conformers of the reference S9 dye differ by about 1 kcal/mol, it is important to probe the conformational dependence of the properties of S9 rotamors (i.e. *cis*-S9 and *trans*-S9). Herein a comparison of the results of such calculations is given.

For *trans* conformation of the reference S9 dye sensitizer, the frontier molecular orbital energy obtained by CPCM-B3LYP/6-311G(d)//CPCM-PBE0/6-311G(d) level of theory are calculated as -4.89 eV, -2.77 eV and 2.12 eV for HOMO, LUMO, and HOMO-LUMO gap, respectively. These calculated values are in excellent agreement with the cis conformer. For example, the HOMO-LUMO energy gap of the trans form differs only by less than 0.05 eV from the calculated values for *cis*-S9 conformer using the same level of theory.

Atomic charges according to the natural bond orbital (NBO) scheme of the π-conjugated bridges of the two conformations are given in Figure S3(a)-(b). As seen from the figure, the atomic NBO charges are almost exactly the same for both conformers.

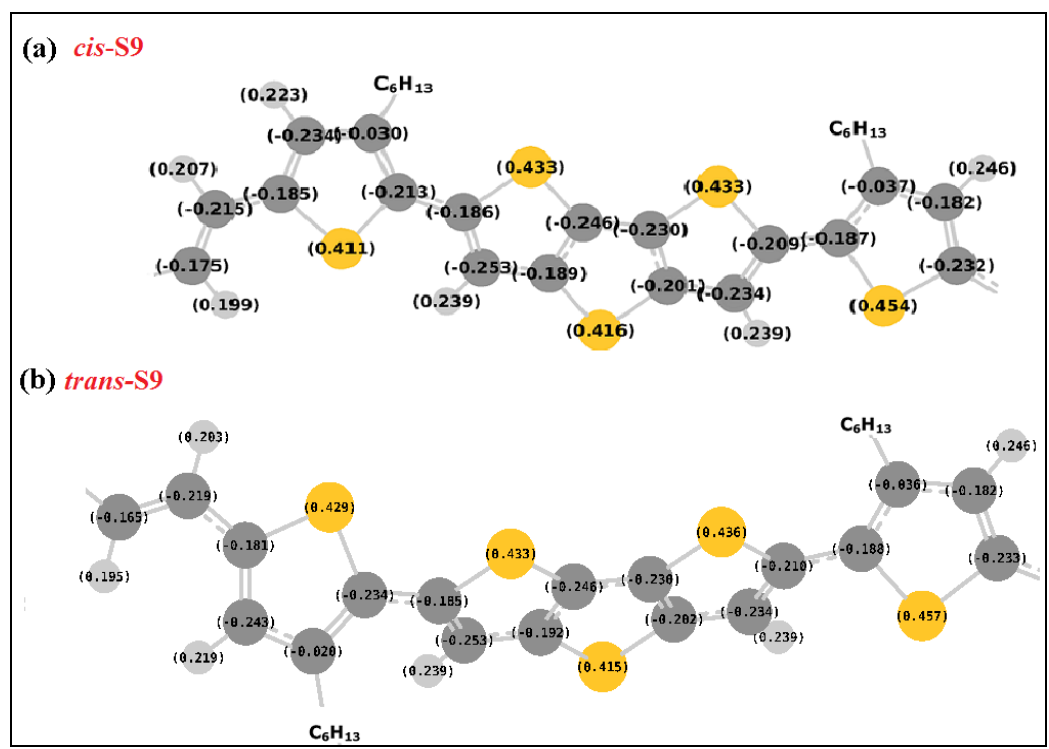

Fig S3: The NBO charge of atoms in the linker of *cis*(a) and *trans*(b) conformers of the reference S9 dye. Note that hexanyl chains are not included.

The $β_{tot}$ are calculated as 805 esu for *cis*-S9, whereas 621 esu for *trans*-S9. It can be seen that the hyperpolarizability of the two rotamors differ by almost 20%. This finding is in agreement with another study on the conformational dependence of the first hyperpolarizability of other conjugated molecules [4]. As suggested in this reference, in such cases one can "consider only one conformation for estimation of hyperpolarizability" [4].

Based on the mean absolute error (MAE) criterion of the BHandH functional (employed to calculate the absorption wavelengths of the three most dominant peaks, Refer to Table S4), the *cis*-S9 gives MAE of 18 nm whereas the *trans*- S9 form produces MAE of 29 nm. This suggests that the TDDFT calculations on the *cis*-S9 are in better agreement with the experimentally measured values compared to those of the trans conformer.